\documentstyle[12pt,aasms4]{article}  

\def\cy{SN~1997cy}
\def\z{SN~1988Z}
\def\bw{SN~1998bw}
\def\d{SN~1997D}
\def\kms{km s$^{-1}$}
\def\Ha{H$\alpha$}
\def\Hb{H$\beta$}

\def\Mv{$M_{\rm V}$}

\def\M{$M_{\odot}$}

\def\m100{mag/100$^d$}
\def\ni{{$^{56}$Ni}}
\def\c57{{$^{57}$Co}\/}
\def\co{{$^{56}$Co}}
\def\fe{{$^{56}$Fe}\/}
\def\ti44{{$^{44}$Ti}\/}
\def\r0{{$R_0$}}
\def\ltsima{$\; \buildrel < \over \sim \;$}
\def\ltsim{\lower.5ex\hbox{\ltsima}}
\def\gtsima{$\; \buildrel > \over \sim \;$}
\def\gtsim{\lower.5ex\hbox{\gtsima}}

\begin{document}
\lefthead{Turatto et al.}
\righthead{Supernova 1997cy}

\title{The properties of Supernova 1997cy\\
associated with GRB~970514
\footnote[0]{Based on observations collected at ESO-La Silla and Paranal}}

\author{M. Turatto\altaffilmark{1}, T. Suzuki\altaffilmark{2,3}, P.A.
Mazzali\altaffilmark{3,6},
S. Benetti\altaffilmark{5}, E. Cappellaro\altaffilmark{1},
\author{I.J. Danziger\altaffilmark{6},
 K. Nomoto\altaffilmark{2,3}, T. Nakamura\altaffilmark{2},
T.R. Young\altaffilmark{2,4}, F. Patat \altaffilmark{7}} }

\altaffiltext{1}{Osservatorio Astronomico di Padova, vicolo
dell'Osservatorio 5, I-35122 Padova, Italy}
\altaffiltext{2}{Department of Astronomy, School of Science, University
of Tokyo, Bunkyo--ku, Tokyo 113-0033, Japan}
\altaffiltext{3}{Research Center for the Early Universe, School of
Science, University of Tokyo, Bunkyo--ku, Tokyo 113-0033, Japan}
\altaffiltext{4}{University of Arizona, Steward Observatory, Tucson, USA}
\altaffiltext{5}{Telescopio Nazionale Galileo,Calle Alvarez de Abreu, 70,
E-38700 Santa Cruz de La Palma, Canary Islands, Spain}
\altaffiltext{6}{Osservatorio Astronomico di Trieste, via G.B.
Tiepolo 11,  I-34131 Trieste, Italy}
\altaffiltext{7}{European Southern Observatory, Alonso de Cordova
3107, Vitacura, Casilla 19001 Santiago 19, Chile}

\bigskip\bigskip

\normalsize

\begin{abstract} 
The extraordinary \cy\/ associated with GRB~970514 has been observed 
photometrically and spectroscopically for nearly 2 yr. At the time of discovery
\cy\/ was the brightest SN ever observed (\Mv$\le-20.1$, v$_{\rm hel}=19140$
\kms, $H_0=65$ \kms Mpc$^{-1}$). Up to the last available observations (600
days after the GRB) the total time-integrated flux was equal to or larger than
that expected from the complete thermalization of the $\gamma$-rays produced by
2.3 \M of \co. However, starting already on day 60 the luminosity decline is 
slower than the \co\ decay rate, indicating that the SN ejecta was interacting
with circumstellar material (CSM). The interaction appeared to weaken around
day 550. The spectra of \cy\/ are dominated at all epochs by \Ha\/ emission,
which shows at least 3 components of different widths, as in \z. Several other
lines with different widths are also visible, especially at early epochs.  The
entire light curve of \cy\/ is reproduced by a model of the interaction of the
very energetic ($E=5 \times 10^{52}$ ergs) ejecta of a massive star (25 \M)
with the CSM, with some contribution from radioactive decays. The CSM could
have been ejected with a mass-loss rate of $\dot{M}\approx4\times10^{-4}$ \M\/
yr$^{-1}$ as the progenitor star evolved from a BSG to a RSG about $10^4$ yr
before the explosion. The lack of oxygen and magnesium lines in the spectra at
nebular phases poses a problem for models requiring high mass progenitors. The
possibility that most of the core material of the progenitor has fallen onto a
massive black hole so that the reverse shock dies at the inner edge of the H/He
envelope is discussed.
\end{abstract}


\section{Introduction}

The spatial and temporal coincidence between the GRB980425 and SN~1998bw
(\cite{gala}, \cite{iwam}) raised the issue of the association of (some) GRBs
with SN explosions.  So far the study of the statistical correlation of SNe
with BATSE and BATSE/Ulysses bursts has given contradictory results. Some
authors have proposed that all GRBs originate from SNe~Ib/c. These are
core-collapse SNe whose progenitors have lost most or all of the hydrogen
envelope (\cite{wang}). However, only a weak correlation was found between the
general list of SNe and GRBs (\cite{kipp}).  As an alternative it has been
suggested that only a small fraction of the GRBs originate from asymmetric
explosions of rare, highly energetic SN Ib/c (\cite{iwam}; \cite{woos}). In
particular, Bloom et al. (1998) on the basis of a model for the radio emission
of SN~1998bw have derived the characteristics which a SN must have in order to
produce a detectable GRB (weak single burst).  Again they indicate highly
energetic Ib/c events as best candidate sources of (some) GRBs.

Germany et al.\ (1999) proposed GRB 970514 as another compelling association
between a highly energetic SN and a GRB. \cy\/ was discovered on July
16, 1997 (\cite{germ_97}) in the compact and faint galaxy n.342 of the cluster
Sersic 40/6 = Abell 3266 (z = 0.059; $\sigma=1211$ \kms; \cite{gree}). The
relatively narrow H$\alpha$ emission led to the classification of \cy\/
as a Type IIn. The epoch of the SN explosion is constrained by a prediscovery
limit on March 12 and is consistent, within the uncertainties, with the
association with GRB 970514, whose $3.7^\circ$ error box was centered only
$0.88^\circ$ away.

The probability of a chance association of the SN with one of the 119 BATSE
events which occurred in the period between the pre-discovery observation and
the discovery images is 0.7\% (\cite{germ}). After \bw, this is the most
compelling case for SN/GRB association. GRB 970514 was a single-peak burst
detected also above 300 keV. At the distance of \cy, the burst energy would be
$\sim 4 \times 10^{48}$ erg, which makes GRB 970514 more energetic than GRB
980425, but orders of magnitude weaker than other bursts with measured
redshifts (\cite{germ}).

In this Letter we present and discuss optical observations of \cy\/ obtained up
to 2.5 years after discovery, together with a model of the light curve derived
with a high progenitor mass and explosion energy.

\section{Light Curve}

Imaging photometry of \cy\/ has been obtained in BVRI with several telescopes
at ESO--La Silla. Typical seeing was 1.2 arcsec.  A sequence of local reference
stars has been measured on seven photometric nights in order to calibrate
non-photometric data. Very deep VLT/FORS1 imaging (seeing 0.6 arcsec) in V and
R about 2.5 yr after discovery did not detect the SN. Although there is very
good agreement for the stars in common with Germany et al. (1999), we note a
difference between our SN magnitudes and their estimates. This is small
(0.15mag) at maximum, but it increases with epoch up to 0.4 mag. We verified
that measuring the SN magnitude with different techniques, PSF-fitting or 
subtraction of pre-discovery images (kindly provided by B. Schmidt, priv.
comm.) - the method used by Germany et al. - produces similar results even at
late epochs. Artificial star experiments show that at all epochs for which we
could make the comparison between the data sets our typical errors are smaller
than 0.1 mag. We believe that the inconsistency is partially due to the
different passbands (\cite{germ}), although the modest spectral evolution of
\cy\/ should not cause a temporal dependence, and/or to incorrect flux
calibration of low signal--to--noise data.  In particular there is the
possibility of contamination by two nearby (2 arcsec) bright knots (SE and SW)
inside the parent galaxy, which could be severe for imaging at small scales or
with bad seeing.

The light curve of \cy\/ does not conform to the classical templates of SN II,
namely Plateau and Linear, but resembles the slow evolution of the Type IIn
\z\/ (\cite{tur93}). Assuming $A_V=0.00$ for the galactic extinction
(\cite{burs}) we get for \cy\/ an absolute magnitude at maximum \Mv$\le-20.1$.
This makes \cy\/ the brightest SN II discovered so far, brighter than the very
luminous SN~1979C and 1983K (M$_V=-19.9$ and $-19.3$, \cite{pata}, with
$H_0=65$ \kms\ Mpc$^{-1}$). Note that Schlegel et al. (1998) give $A_V=0.07$.
As far as we know, no detections at other wavelengths have been reported, 
except for some not very stringent radio limits at 20 cm (\cite{germ}).

\begin{figure*}
\plotone{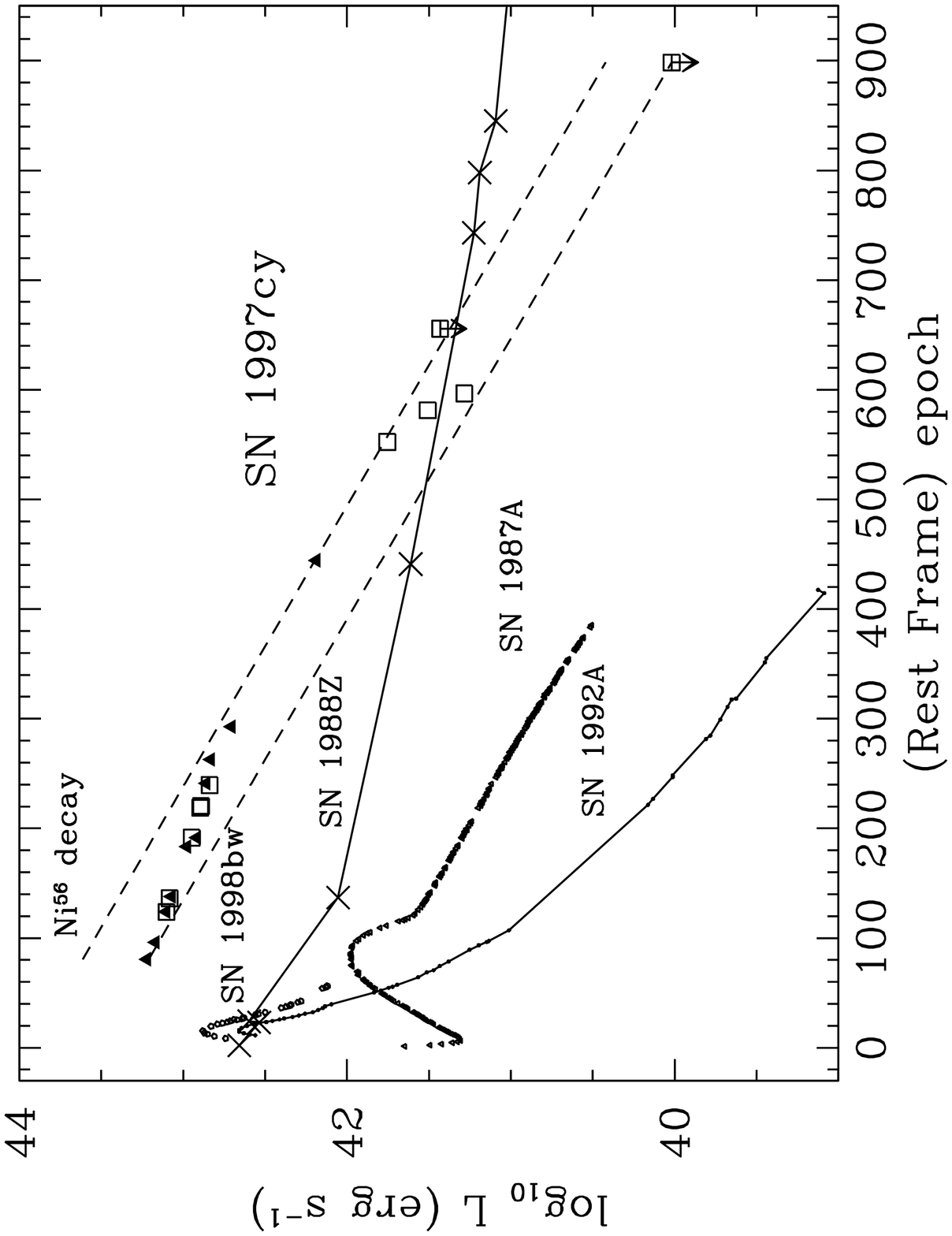}
\figcaption[sn97cy_bolo.eps]{
The {\sl uvoir} (0.32 to 1 $\mu$) bolometric light curve of \cy\/ (using the
luminosity distance, A$_{\rm B}=0$ and M$_{B,\odot}=4.72$). Squares come from
the BVRI photometry and triangles from the spectrophotometry. The dashed lines
indicate the $^{56}$Co decline slope. The rest frame epoch is computed from GRB
970514. Also shown are the bolometric light curves of SNe 1987A, 1988Z (IIn),
1992A (Ia) and 1998bw (Ic pec), associated to GRB 980425. Typical errors are
0.15 dex.
\label{lcbolo}}
\end{figure*}

The {\it uvoir} bolometric light curve of \cy\/ obtained using the available
BVRI photometry and the spectrophotometry is shown in Fig.~\ref{lcbolo}.
Integrating the bolometric light curve we estimate  that the SN has emitted
about $2 \times 10^{50}$ ergs from the day of discovery to our last detection,
not including the GRB luminosity. We stress that the {\it uvoir} determinations
are by necessity lower limits of the true bolometric luminosities.

\section{Radioactive decay vs. Interaction}

The late time light curve of most SNe is powered by the radioactive decay of
\co\/ to \fe. If the hard $\gamma$-rays from the decay are completely
thermalised in the ejecta, the mass of \ni, the parent isotope of \co, can be
estimated directly from the luminosity on the tail of the light curve. Typical
\ni\/ masses for SNe II are $\sim 0.08$ \M\/ (e.g. SN~1987A), but extreme cases
range from 0.002 \M\/ for \d\ (\cite{tur97d}) to 0.3 \M\/ for SN~1992am
(\cite{schm00}).

Between 60 and 120 days after the outburst, which is taken to be coincident
with the GRB, the light curve of \cy\/ matches the decline rate expected if the
energy input was the decay of \co. The estimated initial mass of \ni\ is 2.3\M.
At 4 months the SN decline becomes slower, suggesting that another source of
energy is present. Between days 250 and 550 the light curve again matches
the \co\/ decay of 6\M, which is clearly dubious. Then the light curve suddenly
drops, as does the \Ha\/ emission. On day 655 the SN could not be detected with
the ESO 3.6m telescope. The last points of our bolometric light curve lie above
the radioactive tail of 2.3 \M\/ of \ni, but are compatible with a value close
to that proposed by Germany et al. (1999) if we allow for some interaction.
However, our observations indicate that the late drop mentioned above occurs
about 100 days later than estimated by Germany et al.

Other observable signatures of the interaction, besides the slow decline, are
complex line profiles evolving with time, easily detectable in the prominent
\Ha\/ line, and powerful radio and X-ray emission (\cite{aret}). \Ha\/ in \cy\/
shows an unresolved emission and at least two broader components, which become
narrower with time.  At the epoch of the first observation the broadest
component has FWHM $= 12800$ \kms, and its flux dominates over the intermediate
(FWHM $= 4300$ \kms) one. One year after the explosion, the broadest component
has almost disappeared and the intermediate component (now with FWHM $= 2000$
\kms) is the most evident spectral feature in the spectrum.  \cy\/ was last
detected on April 12, 1999, when only a faint H$\alpha$ emission was detected
at the 3-$\sigma$ level (F$_\lambda=2 \times 10^{-16}$ ergs s$^{-1}$ cm$^{-2}$;
FWHM $= 1000$ \kms).

In Figure 2 we show representative spectra of \cy\/ at various epochs, and in
Figure 3 we compare the spectra of \cy\/ and \z, since a certain spectral 
similarity between the two SNe is possibly suggestive. Lines of different
widths are present in the spectra of \cy. At early epochs there are very broad
bands (FWHM $\sim 300$ A) probably blends of several lines, the strongest of
which are measured at 4150, 4900, 5800, 7800 and 9100 A 
(CaII IR triplet). Narrow lines include \Ha, \Hb, and [OIII] 5007 A.
Intermediate width lines (FWHM $\sim 2500-4300$ \kms) include HeI 5876 A,
which appears in emission between day 100 and 150 (cf. Fig.~3).

A multiwavelength study of \z\/ showed that the energy radiated from radio to
X-rays was as high as $10^{52}$ ergs (Aretxaga et al. 1999), suggesting 
complete reprocessing of the mechanical energy of the ejecta in only a few
years. The behaviour of the interaction between the SN ejecta and a dense
Circumstellar Medium in SN~1988Z has been modelled by Terlevich et al. (1992)
in the framework of the Compact Supernova Remnants (cSNR) scenario, and by
Chugai \& Danziger (1994), who invoked the presence of a rarefied wind together
with a denser component in the form of clumps or an equatorial wind. The
optical light curve, the X--ray emission, the \Ha\/ emission and the line width
evolution of SN~1988Z are generally reproduced by these models but the
resulting ejecta and CSM masses are model dependent. In what follows both the
similarities and differences between SN~1988Z and SN~1997cy should be kept in
mind. The two SNe are similar in apparently having a 3-component profile of the
dominant \Ha\ emission line, and their light curves, though not equally well
sampled, have the same slow decline for the first two years. SN~1988Z was an
order-of-magnitude less luminous than SN~1997cy, while the complex of line
blending in the 4000-6000 A region is clearly different, as shown by careful
inspection of Fig.3.

\section{Modelling the light curve}

The similarities between \cy\/ and \z\/
suggest that we investigate an interaction model for \cy\/ as well. 
Our exploratory model considers the explosion of a massive star of $M = 25
M_\odot$ with a parameterized explosion energy $E$.  We assume that the
collision starts near the stellar radius at a distance $r_1$, where the density
of the CSM is $\rho_1$, and adopt for the CSM a power-law density profile $\rho
\propto r^{n}$.  The parameters $E$, $\rho_1$, and $n$, are constrained from
comparison with the observations. The numerical code and input physics are the
same as developed by Suzuki \& Nomoto (1995).

As in the cSNR model the regions excited by the forward and reverse shock emit
mostly X-rays. The density in the shocked ejecta is so high that the reverse
shock is radiative and a dense cooling shell is formed (e.g. Suzuki \& Nomoto
1995; \cite{terl}).  The X-rays are absorbed by the outer layers and core of
the ejecta, and re-emitted as UV-optical photons. Narrow lines are emitted from
the slowly expanding unshocked CSM photoionized by the SN UV outburst or by the
radiation from the shocks; intermediate width lines come from the shock-heated
CSM; broad lines come from either the cooler region at the interface between
ejecta and CSM, or the unshocked ejecta heated by inward propagating X-rays.

Figure~\ref{lcmod} shows the model light curve which best fits the
observations.  The model parameters are: $E = 5\times10^{52}$ erg, $\rho_1 =
4\times10^{-14}$ g cm$^{-3}$ at $r_1 = 2 \times10^{14}$ cm (which corresponds
to a mass-loss rate of $\dot{M}=4\times10^{-4}$ \M\ yr$^{-1}$ for a wind
velocity of 10 \kms), and $n = -1.6$.  The large mass-loss episode giving rise
to the dense CSM is supposed to occur after the progenitor makes a loop in the
HR diagram from BSG to RSG.  In this model, the mass of the low-velocity CSM is
$\sim 5$ \M, which implies that the transition from BSG to RSG took place about
$10^4$ yr before the SN event.

The large CSM mass and density and the very large explosion energy are
necessary to have large shocked masses and thus to reproduce the observed high
luminosity.  In models with low $E$ and high $\rho_1$ the reverse shock speed
is too low to produce a sufficiently high luminosity. For example, a model with
$E = 10^{52}$ erg and $\rho_1$ as above yields a value of $L_{\rm UVOIR}$ lower
than the observed one by a factor of $\sim$ 5. For high $E$ or low $\rho_1$,
the SN ejecta expand too fast for the cooling shell to absorb enough X-rays to
sustain the luminosity.  Therefore, $E$ and $\dot{M}$ are constrained within a
factor of $\sim$ 3 of the reported values.

The shape of the light curve constrains the circumstellar density structure. 
If $n = -2$, the case of a steady wind, is used, $L_{\rm UVOIR}$ decreases too
rapidly around day 200. The observed light curve shows a first, slight drop
after day $\sim$ 300. This can be reproduced if the CSM density is assumed to
drop abruptly at the radius reached by the forward shock at day 300, so that
the collision becomes weaker afterwards.  This change of the CSM density marks
the transition of the progenitor from BSG to RSG, which should have occurred
$\sim 10^4$ yr before the SN explosion.  This is consistent with the
simultaneous decrease of the H$\alpha$ luminosity mentioned in Sect.4.

After this date, the light curve is powered only by the reverse shock propagating
into the ejecta. After day $\sim$ 550 the observed light curve shows a second,
sharper drop, which is reproduced (Figure~\ref{lcmod}) assuming that after the
reverse shock has propagated through $\sim$ 5 $M_\odot$ of ejecta it encounters
a region of very low density and thus it dies.  In other words, the model
assumes that most of the progenitor's core material fell into a massive ($\sim
10 M_\odot$) black hole, while only the extended H/He envelope of $\sim$ 5
$M_\odot$ was ejected. Therefore, both forward and reverse shocks propagated
inside H/He dominated ejecta.

However, the fact that the ejecta-CSM model with high explosion energy
reproduce the light curve does not necessarily mean that the radioactive energy
input is small, because of the uncertainty in the model fitting.  To constrain
the contribution from the radioactive decay, we calculated the optical light
curve of the SN caused by the combination of shock heating and radioactive
decay. We find that a contribution of up to 0.7 \M\ of $^{56}$Ni is allowed. 
However, the decline rate of the radioactive component after day 60 is much
faster than the half-life of the \co\/ decay, because the explosion is so
energetic that only a fraction of $\gamma$-rays is trapped in the ejecta.  This
implies that radioactive decay cannot explain the slope of the observed light
curve even between 60 and 120 days.

\section{Discussion}

The spectra of \cy\/ obtained within about 100 days from the GRB show broad
emission lines which disappear at later times leaving only a blue continuum.
Broad emission lines originate in the fast expanding ejecta which, in the
interaction model, absorb X-rays produced in the shocks. This is in principle
not very different from the mechanism active in the case of radioactive
powering, i.e. the deposition of hard radiation ($\gamma$-rays in that case).
The width of the broad lines, 13,000 \kms, is consistent with the outer
velocity of the expanding ejecta in the interaction model ($\sim 15,000$ \kms).

Broad Ca~II emission lines are easily identified. However, the Fe~II] and
Fe~III] emission lines strong in the nebular spectra of Type Ia SNe (SNe Ia)
are not obviously identified. This could mean that either the ejected \fe\/
(i.e. \ni) mass is small or that the densities are so high that forbidden lines
are collisionally quenched. In fact the broad feature near 4500 A might be
attributed to the FeII multiplets 37 and 38, while the very broad feature near
5000A may be dominated by a blend of the FeII multiplets 42, 48, and 49.

High densities in the ejecta of \cy\/ appear on the other hand an almost
inevitable conclusion: with a characteristic velocity not much higher than that
of SNe~Ia nebulae, the interaction model for \cy\/ predicts an ejecta mass of
$\sim 5$ \M, i.e. a factor of $\sim 4$ larger. The epoch of the \cy\/ spectra
is also much earlier than it is for SNe~Ia in their nebular epoch (100 v. 300
days), thus adding another factor of $\sim 3^3$ to the density. Thus we can
expect the typical density in the \cy\/ ejecta to be $\sim 100$ times that of
SNe~Ia at 300 days. This implies densities of about $10^8$ cm$^{-3}$. Electron
densities of course depend also on the degree of ionization, but values of at
least $10^7$ cm$^{-3}$ can be expected.

If the forbidden lines are quenched, emission could occur in permitted Fe~II
lines which may be present in the spectrum of \cy, as suggested above.
Nevertheless, the excitation of FeII lines in SNe and AGN's involves processes
not yet fully understood. The H$\alpha$ profile of \z\/ was  explained within
the ejecta--CSM interaction scenario by Chugai \& Danziger (1994), who derived
a mass for the ejecta smaller than 1 \M, with a progenitor main sequence mass
of 8 to 10 \M. Modelling the light curve of \z\/ as the result of interaction
may therefore be worthwhile, and it might indicate a larger mass for \z.
However, another property \cy\ and \z\ have in common is that neither [OI] 6300
nor MgI] 4571 have been detected at any stage in their spectra. The critical
densities of these lines are expected to be a few times 10$^6$ and a few times
10$^9$, respectively. The envelope density at late phases would not be expected
to exceed these values. Since only stars with masses less than 8-10 \M\ (Nomoto
1984) are expected to synthesize little oxygen, one is left with the conundrum
of the missing oxygen and magnesium if it is assumed that these SNe originated
from significantly higher mass stars.

The model described in the previous section might resolve this conundrum: O
and Mg synthesised in the core have fallen into the black hole; the ejecta are
basically the H/He layers and thus contain the original (solar abundance) heavy
elements plus some heavy elements mixed from the core before fall back.  If the
SN formed a black hole, the progenitor mass was probably  larger than $\sim 25
M_\odot$ (e.g. Ergma \& van den Heuvel 1998).

The large energy implied from the modelling of \cy\/ and indeed its high
luminosity alone might suggest a causal connection between
energetic SN events (hypernovae) and the occurrence of GRB's. It appears that
hypernovae occur in a variety of SN types.  Objects such as SNe 1998bw and
1997ef (Iwamoto et al. 2000) are of Type Ic, while \cy\/ is of Type IIn. The
recent SN~1999E (Cappellaro et al. 1999), whose spectrum is identical to that
of \cy, may be another Type IIn hypernova. Objects such as \z\ also deserve
renewed study for the reasons discussed in this paper.

\bigskip

This work has been supported in part by the grant-in-Aid for COE
Scientific Research (07CE2002) of the Ministry of Education, Science,
and Culture in Japan.

\newpage

\begin{figure*}
\plotone{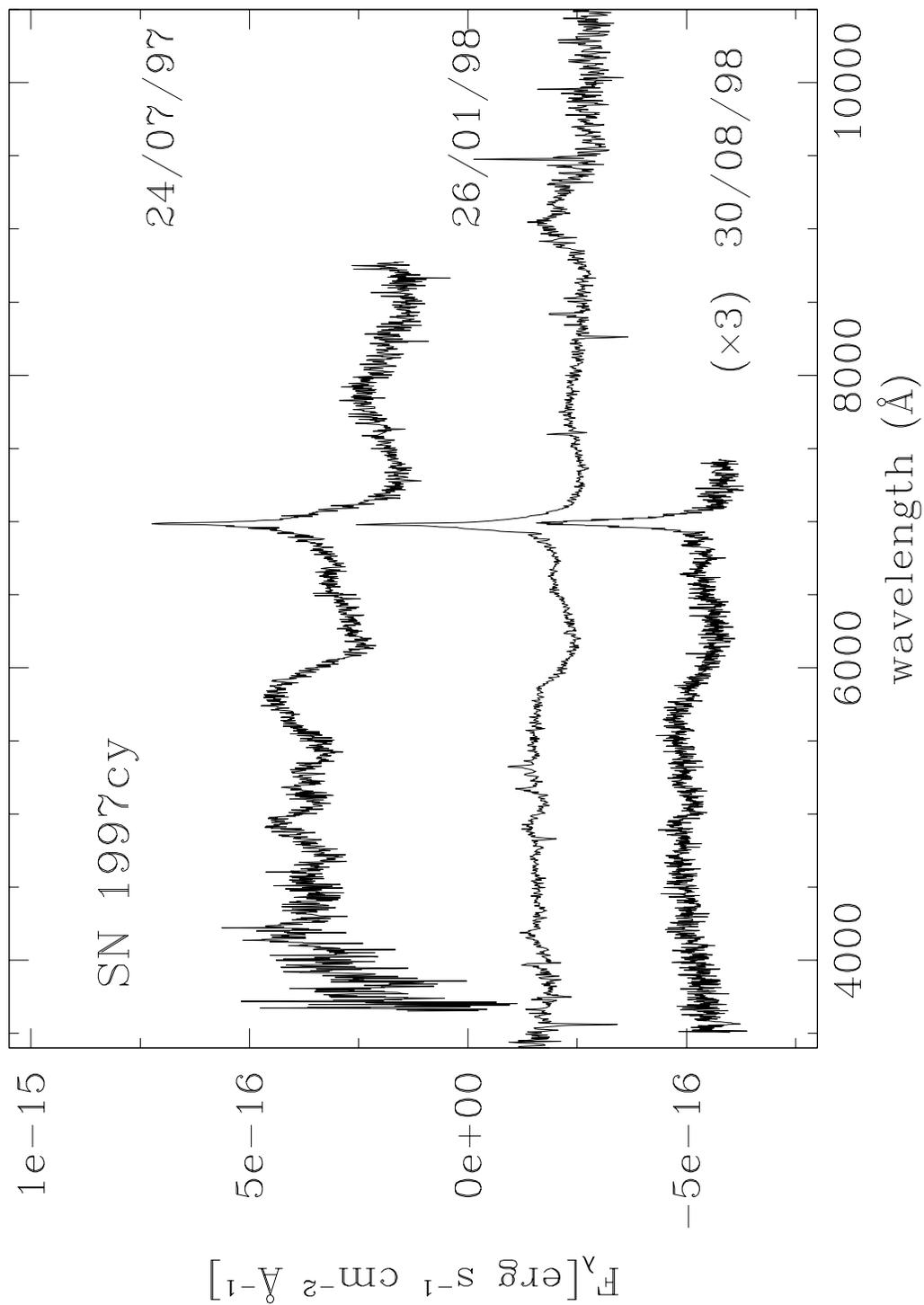}
\figcaption[sn97cy_spev.eps]{ Three spectra describing the evolution of
SN~1997cy. The ordinate axis refers to the first spectrum (top). Other
spectra have been shifted downwards by $3 \times 10^{-16}$ ergs s$^{-1}$
cm$^{-2}$ A$^{-1}$ and the last multiplied by a factor 3.
\label{spev}}
\end{figure*}

\begin{figure*}
\plotone{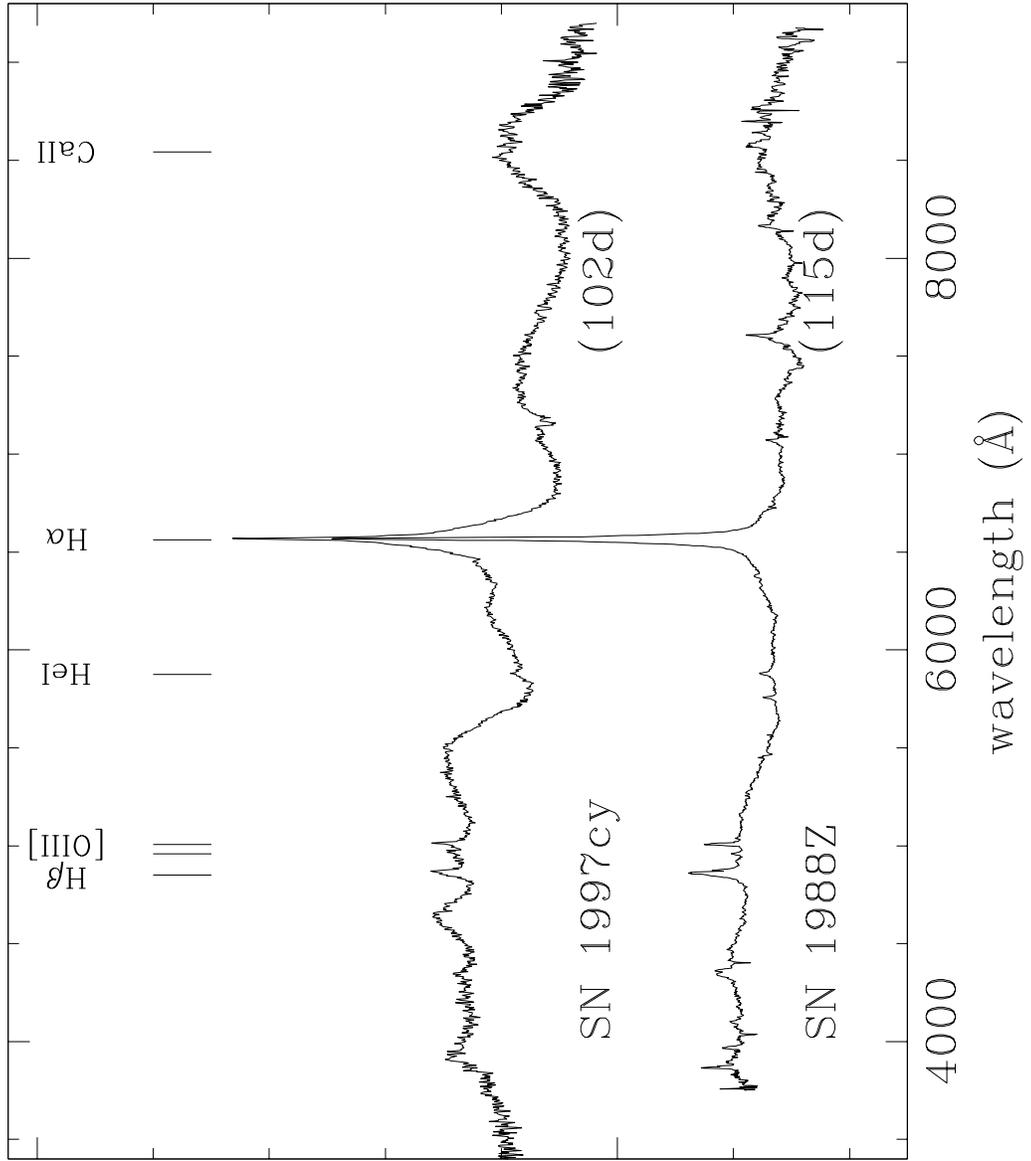}
\figcaption[sn97cy_cfr.eps]{
Comparison of the spectra of SNe~1997cy and 1988Z  (Turatto et al. 1993) at 
comparable epochs. Both objects show broad lines in the blue part as well as a
multicomponent profile of Balmer lines.
\label{spev}}
\end{figure*}

\begin{figure*}
\plotone{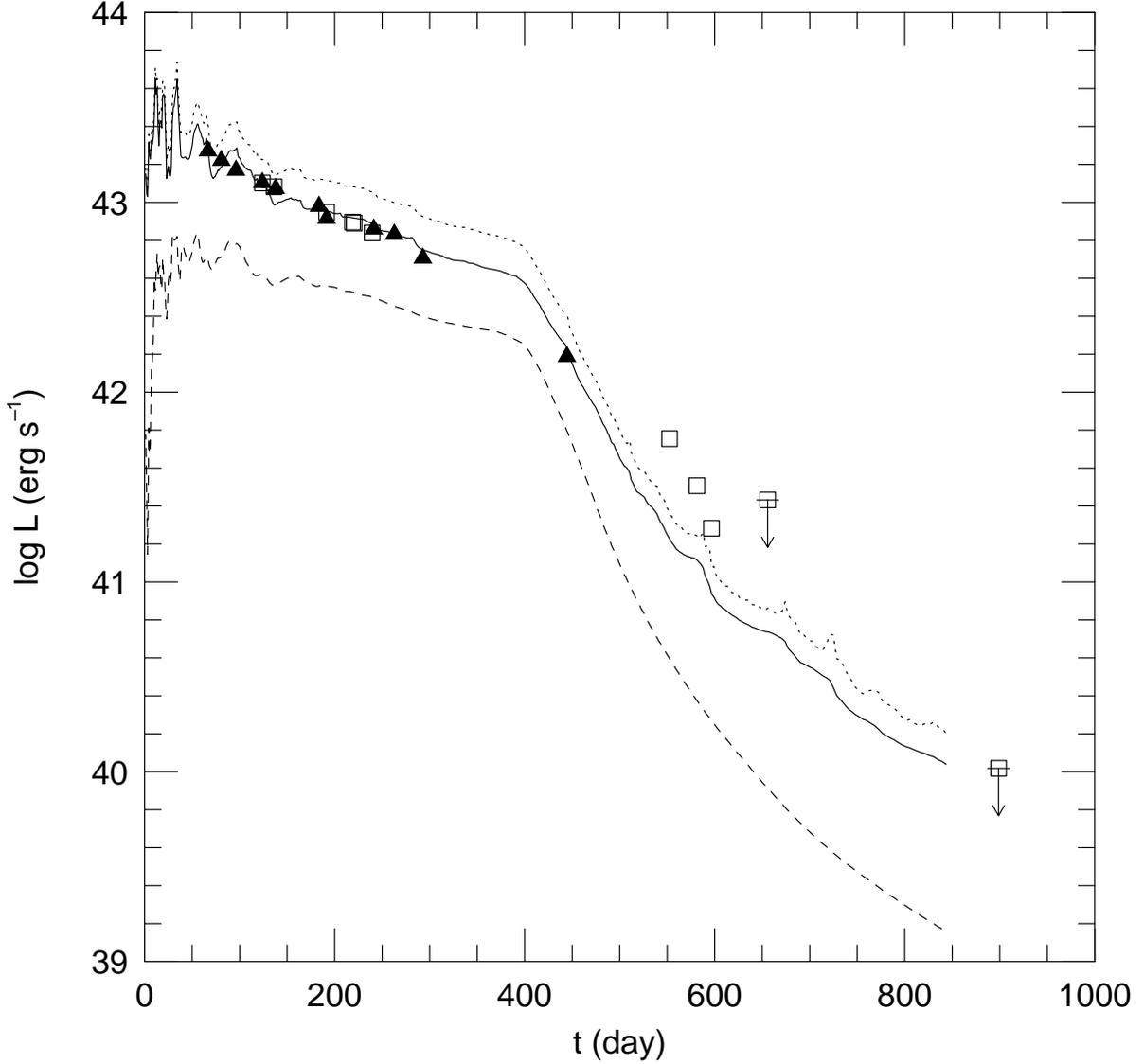}
\figcaption[sn97cy_lcmodfin.eps]{
The synthetic light curve for SN~1997cy obtained with circumstellar interaction
model. The model parameters are: $E = 5\times10^{52}$ erg, $\rho_1 =
4\times10^{-14}$ g cm$^{-3}$ at $r_1 = 2 \times10^{14}$ cm (which corresponds
to a mass-loss rate of $\dot{M}=4\times10^{-4}$ $M_\odot$ yr$^{-1}$ for a wind
velocity of 10 km s$^{-1}$), and $n = - 1.6$.  Shown are the total luminosity
from the shocked ejecta $L_{\rm tot}$ (the dotted curve), the UV-optical
luminosity $L_{\rm UVOIR}$ (solid curve), and the luminosity of the X-rays
escaping from the ejecta $L_{\rm X}$ (dashed curve), where $L_{\rm tot} =
L_{\rm UVOIR} + L_{\rm X}$.
\label{lcmod}}
\end{figure*}

\end{document}